\documentclass[a4paper,conference]{IEEEtran} 
\ifCLASSINFOpdf
\else
\fi
\usepackage[cmex10]{amsmath}
\usepackage{amssymb,verbatim}
\usepackage{graphicx}

\begin{document}
%
\title{%
A Correct Security Evaluation of \\ 
Quantum Key Distribution
}

\author{
\IEEEauthorblockN{Osamu Hirota \\}
\IEEEauthorblockA{
Professor Emeritus \\
 at the Quantum ICT Research Institute, Tamagawa University\\
6-1-1 Tamagawa-gakuen, Machida, Tokyo 194-8610, Japan\\
{\footnotesize\tt E-mail: hirota@lab.tamagawa.ac.jp} \vspace*{-2.64ex}}
}

\maketitle

\begin{abstract}
There is no doubt that quantum key distribution is an excellent 
result as a science. However, this  paper  
presents a view on quantum key distribution (QKD) wherein 
QKD may have a difficulty to provide a sufficient 
security and good communication performance in real world networks. 
In fact, a one-time pad forwarded by QKD model with $\bar{\epsilon}=10^{-6}$ 
may be easily decrypted by key estimation.
Despite that researchers know several criticisms on the theoretical 
incompleteness on the security evaluation, Portmann and Rennner, and others 
still avert from the discussion on criticism, and experimental groups tend to 
make exaggerated claims about their own work by making it seems that QKD is  
applicable to commercial communication systems.
All such claims are based on a misunderstanding of the meaning of 
criteria of information theoretic security in cryptography.
A severe situation has arisen as a result, one that will 
impair a healthy development of quantum information science (QIS).
Thus, the author hopes that this paper will help to stimulate 
discussions on developing a more detailed theory.
\end{abstract}

%
\IEEEpeerreviewmaketitle
\section{{\bf{Introduction}}}
Quantum information science (QIS) may one day bears fruitful applications. 
The author believes this, but the present state of affairs gives 
pause for thought.
That is, many researchers believe quantum key distribution (QKD)
 to be a practical application. 
Theoreticians have performed security analyses on  QKD [1-7], and 
experimentalists have demonstrated many QKD systems.
They claim that QKD will usher in a revolution in  
information technology, because it has information theoretic 
or unconditional security, something which cannot be realized by classical 
information technology. Moreover, they have represented their 
achievements to the outside world in a way that the physics community 
can now provide the ultimate level of protection against cyber attacks. 

Such prognostications and assertions are not the typical province 
of the physics community (at least, those who are not specialists 
in the actual technology). So they should not take such an action, 
 because their behavior may stimulate unreasonable 
expectations in the general public. In fact, there is still a great 
gap between the scientific issues and its  applications. 

Over the last ten years or so, extensive investigations have 
demonstrated the theoretical incompleteness of QKD security [8,9].\\ 
The main result is as follows:\\
(a) One needs to employ a correct evaluation for guarantee of information 
theoretic security.\\
(b) Even the QKD system is information theoretically secure, 
it cannot provide a sufficiently uniform random key bits 
for application to a one-time pad. \\
(c) So a one-time pad with a generated key by the QKD is not secure. 
 \\

While the criticisms posed by these studies are very reasonable, 
it seems that QKD researchers have ignored them. Yet as a community 
that belongs to the greater physics community, they have 
a responsibility to make the truth clear to the public, 
because it is the way of science. 
Unfortunately the present author does not feel that this issue has 
yet to be dealt with, because their action is  as follows:\\
``QKD researchers just neglect the interpretation such as 
failure probability 
in their papers without a correction after the criticism, and 
they  do not care the meaning of $\epsilon$-security, despite they claim that 
this interpretation gives a degree of security in a real system."\\

In fact, if the QKD community accepts the criticism, the present QKD becomes 
meaningless in a practical use.
Thus, the present paper outlines the problems with 
security  claims about QKD, as has been pointed out by a number 
of authors before.

\section{{\bf{Misconception on the security of a one-time pad}}}
 Researchers of experiment claim that a one-time pad is secure. 
However, this is not true, because its security depends strongly on the 
randomness of the key sequence. If the uniformity of the key sequence 
in a one-time pad is not sufficient, its key sequence can be 
easily estimated with a known plaintext attack. 
That is, an attacker can use ``the next bit prediction property described by
 $P(K_{i+1}|K_i)$".
 Note that if the key sequence has perfect uniformity, the one-time pad provides  
so called perfect secrecy or unconditional security, 
because nobody can predict the key sequences. 
Thus, to enjoy the perfect secrecy, one needs a perfectly uniform 
random key sequence or one that is nearly perfectly uniform.

On one hand, researchers of theory assert that QKD can provide
 such a uniform key sequence with high probability (nearly 1),
 and they make this claim on the basis of the trace distance formula.
 Other researchers in cryptography, however, would dispute that one  
  can generate and share a perfectly uniform random sequence 
by using a physical process such as communication process disturbed 
by attackers, 
because any physical process with only one access to a generation source
of a random sequence fails with probability 1 to get perfectly 
uniform randomness,  as is discussed in section V. 
In the following sections, I would like to point out that 
the researchers of QKD  are claiming that they can realize
 physically impossible functions such as generating a perfectly 
uniform long key bits by a communication protocol.

\section{{\bf{Standard security theory of QKD}}}
\subsection{{\bf {Requirements to guarantee security}}}

The primitives of security for QKD are as follows:\\
(a) The laws of quantum mechanics including Heisenberg uncertainty 
principle has the consequence that any measurement by Eve on 
a channel causes large errors.\\
(b) Security cannot be guaranteed only by detecting errors caused by 
Eve's measurement. Thus, the phrase 
 ``the Eve's error is sufficiently large" is not guarantee of security 
in the cryptographic sense. The degree of  uniformity of the generated key
 has to be quantified in terms of security measures.\\
(c) A one-time pad forwarded by QKD has to be guaranteed to be secure 
against a known plaintext attack(KPA)\\
\\
All researchers in cryptography know of the above primitives.
However, many articles in physics ignore the serious 
issues of (b) and (c), especially the theories of Gottesman [10] and so on [1], 
which focus on only physical layer like (a). They do not consider 
problems from the viewpoint of the user of cryptography 
which are the most important ones affecting practical use.
In the time since the above deficiencies were brought to their attention,
the QKD community has tried to incorporate the issues 
of (b) and (c) in their discussions[2], though it is not sufficient. 
Thus many mathematical techniques in cryptography have been employed to improve 
a security performance. 
Because of that, no one can claim any longer that ``the security of QKD
 is entirely ensured only by physical law".\\
The physics issues are still what the literature stresses [1].
To see where this has taken us, let us visit the standard theory 
of security in QKD  and check how it stands up against requirements 
(b) and (c).

\subsection{{\bf{Definition of security}}}
In the standard theory of QKD, it is claimed 
that the trace distance ($d_{QKD}$) guarantees  the universal 
composability of QKD security. In this section, the author will try to 
make clear what the main issue is in the claim of the unconditional 
security of QKD theory. 

In the formulation of QKD theory by R.Renner,
the claim that the generated key sequence provides unconditional security
 is made by invoking the trace distance criterion ($d_{QKD}$) as follows [2-7]:
\\
\\
{\bf Definition 1} \\
The trace distance is given as follows:
\begin{eqnarray}
d_{QKD} \equiv \frac{1}{2}||\rho_{KE}-\rho_U \otimes \rho_E||_1
\end{eqnarray}
where $\rho_{KE}$ is the density operator of the shared key 
 at the final stage of the protocol,
and $\rho_U$ is that of a uniformly random situation. 
When 
\begin{equation}
d_{QKD} \le \epsilon
\end{equation}
the generated key is called ``$\epsilon$-secure". 
\\

\subsection{{\bf{Operational meaning of the security measure}}}
In order to give an operational meaning to the trace distance, 
 Renner and his group claim that a uniform random key
 can be generated with  probability $(1-d_{QKD})$,  
and the failure probability is $d_{QKD}$. 
If one puts the upper bound of $d_{QKD}$ as $\epsilon$, the probabilities of 
 success and failure of the protocol are (1- $\epsilon$) and 
$\epsilon$, respectively.
In some cases, they take the trace distance to have 
the meaning of the failure probability of the indistinguishability between 
the generated key sequence and a perfectly random sequence. 
Consequently, they claim that the operational meaning of the trace distance is 
a``failure probability".

Thus, they insist that the generated key sequence is always 
a perfectly uniform random bit sequence whenever the protocol succeeds. 
However, they have not given the mathematical reasoning behind their assertion,
and instead refer to its similarity with the classical information
theoretic security model.

Let us cite the origin of these claim. 
The QKD researchers employ the following statistical distance formulation.
\begin{equation}
\delta(P,Q)=\frac{1}{2}\sum_{x \in {\bf X}}|P(x)-Q(x)|
\end{equation}
which can be related with $d_{QKD}$.
Renner relies on the following Lemma in his paper to justify his 
own interpretation ($d_{QKD}$ gives a failure probability).
\\
\\
{\bf (Lemma cited by Renner [2])} \\
Let $P$ and $Q$ be two probability distributions. Then there ``${\bf exists}$" 
a joint probability distribution $P_{{\bf XY}}$, 
such that $P_{{\bf X}}=P$, $Q_{{\bf Y}}=Q$, 
and 
\begin{equation}
\delta(P,Q)=P_r[{\bf X} \ne {\bf Y}]
\end{equation}
\\
Though the lemma itself is correct, Renner has started to stray off course, 
by interpreting it as a statement of the failure probability. 
H.P.Yuen pointed out in his paper[8,9] that the above statement is wrong,
and that it leads to a  misconception about the security analysis of QKD.
In the following, I will detail Yuen's claim.
\\

\section{{\bf{Kato's analysis of the coupling theorem}} }
One of the main problems with the security of QKD is 
the operational meaning of the trace distance or statistical distance.
If one wants to give this security measure a certain operational 
meaning in information theoretic
 security, one has to express it  as a probability 
 for a certain event occurring.
 
A theorem that treats the relation between the statistical distance and 
the probability of an event is called the coupling lemma, but in this paper
 I will call it ``the coupling theorem"". Here I will introduce 
the coupling theorem and its application to QKD 
as discussed by K.Kato [11].

\subsection{{\bf{Statistical distance in classical security theory}}}
Let us revisit the formulation of information theoretics security in
the classical theory.
In Shannon theory [12], a scheme is called  perfectly secure when 
\begin{equation}
P(X|C)=P(X) \quad \forall C
\end{equation}
which corresponds to $I(X,C)=0$. This means that $C$ and $X$ are 
 statistically independent. However, to realize such a situation, it requires  
that the key sequence ${\bf{K}}$ has to have perfectly uniform randomness.
When a key sequence is not uniform, one has $I(X,C)>0$. But such a 
mutual information does not provide an operation meaning of a security 
when it is not zero (Note that the mutual information between Alice 
and Eve in the QKD model also does not), because Shannon's information measure 
is not ``information" in the sense of cryptography.
So some researchers have tried to employ the following measure.\\
\\
{\bf Definition 2}\\
Let $X\in{\bf X}$ , $C\in{\bf C}$ , and $K\in{\bf K}$ be the 
message, ciphertext,and running key, respectively.
When it satisfies
\begin{equation}
\delta(P_{\bf XC},P_{\bf X} P_{\bf C}) \le \epsilon
\end{equation}
it is also called $\epsilon$-secure.\\
\\
This gives a measure of a closeness of joint distributions 
$P(X)P(C|X)$ and $P(X)P(C)$. When $\epsilon$ is not zero, 
one again encounters a problem of the operational meaning of 
the quantitative value of $\epsilon$. To make it clear, one needs 
a careful consideration of the coupling theorem. 
So far, this brought a serious confusion.\\
\\
Let us here examine a relation between a statistical 
distance and failure probability for certain binary events. 
Renner has relied on the coupling theorem. 
The exact description is as follows:\\

{\bf Theorem 1} (Coupling theorem)
\\
Let ${\bf X}$ and ${\bf Y}$ be random variables associated with
 two distribution $P_{\bf X}$ and $Q_{\bf Y}$ on a finite set. Then 
 we have
\begin {equation}
 \delta(P_{\bf X},Q_{\bf Y} )\leq  P_r({\bf X} \ne {\bf Y})
\end{equation}
Or there exists a joint distribution $P_{XY}$ 
such that $P_{{\bf X}}=P$, $Q_{{\bf Y}}=Q$, 
and 
\begin {equation}
 \delta(P_{\bf X},Q_{\bf Y} )= P_r({\bf X} \ne {\bf Y})
\end{equation}

Unfortunately, so far there was no theory to make the operational meaning clear.
In spite of this fact, Renner used the second part of the above theorem 
to justify his own interpretation of the key distribution model. 
From the word of the ${\bf ``exist"}$ in the lemma cited by Renner, 
nobody can claim that the mere existence of $P_{XY}$ enables  the statistical
 distance to be interpreted as the failure probability. 
Even if it exists, it corresponds to an unacceptable situation
 for QKD that can be checked very easily as follows:

From the coupling theorem one can make the following case.
 Let ${\bf X}$ be a random variable associated with a distribution
 $P_{\bf X}$ on a finite set. One makes a copy of ${\bf X}$, creating 
 a new random variable ${\tilde {\bf X}}={\bf X}$. This copying process tacitly  
 implies a noiseless channel. On the other hand, let us consider that 
 the random variable ${\bf X}$ is transmitted through a noisy channel, and 
 let ${\bf Y}$ be the random variable at the channel output.
 How close is the initial perfectly correlated pair:
$({\bf X},{\tilde {\bf X}})$ to the noisy channel pair:
$({\bf X},{\bf Y})$. In this case, one has the special property pointed out 
by Nielsen et al [13]. That is, 
 \begin {equation}
 \delta(P_{{\bf X},{\tilde {\bf X}}}, P_{{\bf X},{\bf Y}})= 
P_r({\bf X} \ne {\bf Y})
\end{equation}
Thus, \\
$\quad$``The statistical distance gives a failure probability". \\
But this is a statement about the closeness of a joint probability distribution
 of an imperfect correlation between two random variables and 
that of a perfect correlation which is a copy.  
 Clearly this is not a general setting for a cryptographic model.
Newcomers to QKD sometimes take the above sentence as justification of the 
failure probability interpretation without taking into the feature 
of the cryptographic model account.
\\
In the cryptography model, we have to deal with  the distance 
between the joint distribution of an imperfect correlation and 
that of perfectly independent variables.
The problem is whether one can get the failure probability 
interpretation when the system corresponds to the cryptographic case. 
The answer is no, because all random variables are different 
and one has to take into account the general case in the coupling theorem.

 \subsection{{\bf{From trace distance to statistical distance}}}
 The trace distance is defined in the first stage of the 
 security analysis of QKD. Here let  $\rho$ and $\sigma$ be  density 
operators. Suppose that one applies the same measurement procedure 
to $\rho$ and $\sigma$,
let $P_{\bf X}$, and $Q_{\bf Y}$ be the probability distributions.
Accordingly one has the following relation.
\begin{equation}
 d(\rho, \sigma) =\frac{1}{2}||\rho - \sigma || \ge \delta(P_{\bf X},Q_{\bf Y} ) 
 \end{equation}

Let us recall the trace distance in the QKD model.
\begin{eqnarray}
d_{QKD} \equiv \frac{1}{2}||\rho_{KE}-\rho_U \otimes \rho_E||_1 \nonumber
\end{eqnarray}
The joint probability distributions for  $\rho_{KE}$, and 
$\rho_U \otimes \rho_E$ are $P(K,E)$ and $U(K_U)\times P(E')$, 
respectively. $U$ means an uniform distribution. The statistical distance 
that corresponds to the above $d_{QKD}$ is
\begin {equation}
 \delta =\delta(P_{{\bf K},{\bf E}}, U_{{\bf K_U},{\bf E'}})
\end{equation}
According to the coupling theorem, the statistical distance is upper bounded by 
the failure probability $P_r({\bf K} \ne {\bf K_U})$. 
\begin {equation}
 \delta \leq P_r({\bf K} \ne {\bf K_U})
\end{equation}
If the trace distance is upper bounded by $\epsilon$ and $\epsilon$ is 
the failure probability, one has to conclude the following relation [11]
\begin{equation}
d_{QKD} \ge \delta \quad \leq \epsilon =P_r({\bf K} \ne {\bf K_U})
\end{equation}
It is clear that the above relation has a mathematical contradiction, 
and nobody can claim a probabilistic meaning based on the coupling theorem 
(a general description of lemma cited by Renner)in the QKD model. 
Consequently, $\epsilon$-security of QKD has no 
interpretation of ``failure probability"of a generation of uniform random 
key sequence. 
\\

\subsection{{\bf{Another misuse of the interpretation}}}
M.Koashi published a paper titled 
"Simple security proof of quantum key distribution based on complementarity"
in the New Journal of Physics (2009)[14].
This paper however contains another typical misuse of the meaning of 
the trace distance and fidelity. 

It claims that the trace distance is bounded by a fidelity, and the fidelity 
has an operational meaning as a probability for certain event.
A fidelity is a generalization of an inner product between two density operators, 
and its meaning is a closeness. But Koashi says that a general fidelity 
between two density operators can has an interpretation as  
a probability, because the measurement probability by fidelity form 
between density operator of physical quantity and POVM 
$\Pi=|\varphi_M><\varphi_M|$means a probability.
This reasoning has been definitely described in his book.
That is, his reasoning is a similarity as the following.
\begin{equation}
Tr\{\rho_1 \rho _2\} \quad vs \quad Tr\{\rho_1 \Pi\}
\end{equation}

\section{{\bf{Real security of one-time pad by QKD}}}
So far, an evaluation in the sense of information theoretic security has been 
dealt with mutual information and related function. Reason of why is 
that authors are only interested in a conceptual security apart from 
a computational based security, not operational security meaning. 
A one-time pad forwarded by QKD indeed needs an operational meaning of 
the security measure. Thus, one needs to compare the following proposals.

\subsection{{\bf {From Renner to Yuen}}}
Here I would like to compare the formalisms of the security of one-time pad 
forwarded by QKD, discussed by Renner and Yuen. \\
\\
(a) ${\bf {Renner}}$: One can define the security of one-time pad by 
the trace distance between a real density operator for shared key sequence and 
that of an ideal situation, because the trace distance or its upper bound
 have a meaning of the  probability that the shared key sequence 
does not have a perfectly uniform randomness. 
This is called ``Composability".
\begin{eqnarray}
d_{QKD} \equiv \frac{1}{2}||\rho_{KE}-\rho_U \otimes \rho_E||_1 \leq 
\epsilon_{R}
\end{eqnarray}
\\
(b) ${\bf {Yuen}}$: The trace distance has no interpretation of 
failure probability. 
Then, since one-time pad is a kind of stream cipher, 
one has to evaluate the security by a next bit prediction or 
related concept like a standard theory of cryptography. 
So the one-time pad forwarded by QKD has to be directly evaluated by 
\begin{equation}
P({\bf{K}}|{\bf{Y}}_E) = 2^{-H_{min}({\bf{K}}|{\bf{Y}}_E)} \leq \epsilon_{Y}
\end{equation}
where ${\bf Y_E}$ is the measurement data by Eve's optimum POVM, $H_{min}$ is 
a min entropy.\\
\subsection{{\bf{A correct theory of security evaluation of QKD}}}
Almost all theoretical discussions of the security of QKD have
 dealt with only the physical errors of Eve. They say in effect that 
``the errors of Eve are sufficiently large, so the system is  secure". 
But in cryptography, one has to ensure security for  
the encryption of the data. Let us check out the story of  
 the physical process of QKD and its application to a data encryption
 such as a one-time pad.
Consider a hybrid cipher consisting of a one-time pad and QKD.
How can we realize  perfect secrecy with this hybrid cipher ?
A one-time pad is a perfectly secure when and only when 
 a key sequence of the same length of the plaintext (message) is 
a perfectly uniform random number. That is, 
Eve's estimation probability $P({\bf K}|{\bf Y_E}) $for the key sequence 
before one-time pad encryption is 
\begin{equation}
P({\bf K}|{\bf Y_E}) = 2^{-|{\bf K}|}
\end{equation}
where ${\bf Y_E}$ is the measurement data of Eve, $\Pi_{K}$ is a optimum 
POVM, and $|{\bf K}|$ is the key length.

The security evaluation in the present theory relies on the trace distance,
i.e., $\epsilon$-security. The point is ``What meaning does $\epsilon$ have ?".
It is not the failure probability, but  is related with the uniformity 
of the key sequence. That is, it gives an upper bound for Eve's
estimation probability of the generated key sequence as follows:
\\
\\
{\bf Theorem 2} [9] \\
Let $<d_{QKD}>={\bar{\epsilon} }$ be the averaged trace distance with 
respect to a random error correcting code EC and privacy amplification code
PA. The upper bound of the  averaged estimation probability  of 
the key sequence is given by 
\begin{equation}
<P({\bf K} | {\bf Y}_E)> \quad \le \quad {\bar{\epsilon}}  + 2^{-|{\bf K}|}
\end{equation}
where $|{\bf K}|$ is the key length.\\
\\
The above theorem says that the trace distance and its upper bound give 
an upper bound of the estimation probability of whole key sequence, and also 
asserts that the estimation probability corresponds to the degree of 
uniformity of the key sequence.
If the estimation probability is very large in comparison with 
a perfectly uniform, any security analyst knows of many known plaintext 
attacks against the one-time pad using such a non-uniform key sequence.
So it may be very weak in comparison with 
 computationally secure encryption even though it has information 
theoretic security described by 
$<P({\bf K} | {\bf Y}_E)> \sim {\bar{\epsilon}}$, 
because a one-time pad has no complex algorithm to encrypt the data sequence, 
and algorithm is not necessary to break it. I will show a meaning of the above 
in the following.

\subsection{{\bf{Security of ciphertext only attack }}}

Let ${\bf {C}}={\bf{X}}\oplus {\bf {K}}$ be the ciphertext of 
a one-time pad by the generated key of QKD.
Eve can obtain the exact ciphertext, so her estimation probability can be 
translated into 
\begin{equation}
P({\bf K} | {\bf Y}_E)=P({\bf K} | {\bf C})
\end{equation}

The QKD researchers give a bound of $d_{QKD}$ as the average over 
the random EC and PA. So one should use the Markov inequality 
to converts the averaged evaluation to an individual one. Thus we have 
\\
\\
{\bf Theorem 3} [9]\\
Let us assume the averaged $\epsilon$-security as follows:\\
$<d_{QKD}> \quad \le \quad {{\bar{\epsilon}}}$. 
After application of the Markov inequality two times, one gets 
\begin{equation}
P({\bf K} | {\bf C}) \le {\bar {\epsilon}}^{1/3}  + 2^{-|{\bf K}|} 
\end{equation}
\\
\\
Let us consider an example. Here we assume that 
\begin{equation}
<d_{QKD}> \quad \le \quad {\bar {\epsilon}} =10^{-6}
\end{equation}
From the theorem 2, one has
\begin{equation}
<P({\bf K} | {\bf C})> \quad \le \quad 10^{-6} + 2^{-{\bf |K|}}
\end{equation}
When the length of the generated key is $|{\bf K}|=10^4$, 
the security requirement of the key estimation probability is 
 order of $10^{-3000}$.   That is, 
\begin{equation}
10^{-3000}<< 10^{-6}
\end{equation}
Thus $10^{-6}$ is excessively large 
and it does not give the sufficient security guarantee.

From the theorem 3, the worst case is as follows:
\begin{equation}
P({\bf K} | {\bf C}) \le \epsilon=10^{-2}
\end{equation}
where $\epsilon = {\bar{\epsilon}}^{1/3}$.
This means that $10^4$ bits key sequence may be estimated with the probability 
$\sim 1/100$\\.
\\
On the other hand, in another point of view, 
one can understand the following property.
One bit may be leaked for every $f$ bits in 
$|{\bf{K}}|=l$ generated key bits, wherein $f$ is
\begin{equation}
f= \log_2 \frac{1}{\epsilon}
\end{equation}
That is, the following key bits may be leaked per protocol:
\begin{equation}
|{\bf{K}}|_{leak}=\frac{l}{f}=\frac{l}{\log (1/l)}
\end{equation}
When $|{\bf{K}}|=l=10^4$, and ${\bar{\epsilon}}=10^{-6}$
 (the best experimental result),  
it means $\epsilon=10^{-2}$. So about 1,500 bits per 10,000 bits may be leaked.

\subsection{{\bf{Security of known plaintext attack }}}

One-time pad is a stream cipher, so one needs to check a property of 
next bit prediction of the key sequence.

I will discuss here the real meaning of the quantitative security guaranteed by 
$\epsilon$-security.
Let ${\bf K}$ be the generated key sequence, 
and let ${\bf K}_{(KP)}$ and ${\bf K}_{(Re)}$ be known key sequences in 
the generated key from some known plaintext and the remaining key sequence,
respectively.

From the cryptography theory of stream cipher, one has to consider 
the next bit prediction property. This is just a procedure that 
one tries to estimate the remaining key sequence ${\bf K}_{(Re)}$ 
from the knowledge of ${\bf K}_{(KP)}$. 
\\
\\
{\bf Theorem 4} [15] \\
Let us consider a one-time pad by an imperfect random key 
sequence generated with $\epsilon$-security, where $\epsilon\equiv 2^{-m}$.
When the known keys $|{\bf K}|_{(KP)}=m$ bits, there exists the next bit 
sequence prediction property as follows:
\begin{equation}
P({\bf K}_{(Re)}| {\bf K}_{(KP)})\sim 1 
\end{equation}
\\
\\
Since a one-time pad has no computational complexity, the remaining part
of the key and plaintext may be instantly exposed  by any determined
 high school student. 

Consequently, generated key sequence by QKD does not have sufficient security
 if $\epsilon$ is not on the order of $2^{-|{\bf K}|_{(Re)}}$.
If $\epsilon$ in a real QKD is one the order of $10^{-6} \sim 10^{-14}$, 
 the one-time pad forwarded by such QKD is completely insecure.\\

\section{{\bf{Limitation of security}}}
In this section, we discuss limitations of security in a real QKD.

\subsection{{\bf{Trade-off between $\epsilon$ and key rate}}} 

Let us show that even with the most favorable treatment, the present 
QKD scheme comes to naught as a security guarantee.
Many experimental studies on QKD discuss only 
the key generation rate $r$ and do not indicate the value of ${\bar{\epsilon}}$.
According to  security theory, they have to show how much ${\bar{\epsilon}}$ 
is realized for their own key rate, because there is a trade-off between 
the key rate and ${\bar{\epsilon}}$. Tamaki and Tsurumaru claimed in the 
QIT workshop that one can set ${\bar{\epsilon}}$ arbitrarily. 
But in making this claim, they ignored the following fact. Tomamichel et al. [6] 
showed the trade-off between ${\bar{\epsilon}}$ and key rate. 
An ${\bar{\epsilon}} $-secure key can be extracted out of the 
reconciled key of length, 
\begin{equation}
l({\bar{\epsilon}}) \le n(1-h(Q+\mu))-Leak_{EC}-
\log\frac{2P_{fail}}{\bar{\epsilon}^2\epsilon_{cor}}
\end{equation} 
where $n$ is the block length.
This is a function of ${\bar{\epsilon}}$.
Let us check the property of Eq(26). They fixed the  
security rate as
 \begin{equation}
 S=\frac{{\bar{\epsilon}}}{l}=10^{-14}
 \end{equation}
Then, they gave  numerical examples for the key rate.
When the block length is from $10^{7}$ to $10^4$, the rate is 
$10^{-1} \sim 10^{-2}$. This means ${\bar{\epsilon}} = 10^{-8} \sim 10^{-12}$.
As a result, the best security parameter ${\bar{\epsilon}}$ is one with  
a vanishing key rate as follows: 
\begin{eqnarray}
<d_{QKD}> &\le& {\bar{\epsilon}} = 10^{-14} \nonumber \\
r &=&\frac{l}{n} \sim 0
\end{eqnarray}
This fact was pointed out by Yuen and Kanter.
Thus, experimentalists have to show  both parameters ($\bar{\epsilon}$, $r$),
otherwise the experimental demonstration has no meaning
 in the sense of cryptography.

Consequently, the best averaged $\epsilon$-security of QKD system even with 
sophisticated devices is $10^{-14}$ for 
a block length of $10^{4}$ bits under the zero rate.
If one expects the secure one time pad based on these keys, 
one needs $\epsilon \sim 10^{-3000}$. This is impossible in a real setting.
So  QKD is not practical at all.

\subsection{{\bf{Limitation of privacy amplification}}}
In general, QKD researchers claim that a privacy amplification based on 
Hash function is a key technology to enhance a security of shifted key. 
A role of the Hash function is to reduce a length of generated key for 
making a uniformity in bit sequence of the generated key. However,
since one has to open what kind of Hash function is used, 
``knowledge" of Eve on the bit sequence before the privacy amplification 
is not reduced. Thus, one has to accept the following. \\
\\
{\bf Theorem 5}\\
Any privacy amplification does not improve $P({\bf K} | {\bf Y}_E) $ which 
is fixed at shifted key phase.\\
\\
The above is clear from information causality.

\subsection{{\bf{Physical limitation of uniformity 
in random number generation}}}

Let us discuss the interpretation problem  from the experimental point of view.
The QKD researcher claims that they can realize, by the sophisticated technology, 
average of $\epsilon$ with respect to random EC and random PA as follows:
\begin{equation}
{\bar{\epsilon}} = 10^{-6} \sim 10^{-14}
\end{equation}
for the total generated key $|{\bf{K}}|=10^4$ bits. 
They insist that an upper bound of the trace distance ${\bar{\epsilon}}$ 
is the failure probability for getting the uniform random variable 
and they say that always the uniformity of the generated key is ensured. \\
However, Yuen has repeatedly pointed out that, from the physical point of view, 
the failure probability for getting a perfectly 
uniform random variable is always one [9]. That is, \\
\\
{\bf Remark:} \\
When the trace distance $d_{QKD}$ is not zero, the real failure probability to 
get a perfectly uniform key sequence is nearly 1, and the failure probability
 is practically independent of the value of the trace distance
 (or its classical correspondence:statistical distance).
\\
\\
Iwakoshi reported an experimental study in the QIT workshop on the relation 
between the statistical distance and the uniformity of 
a sophisticated physical random number generator[16]. 
His results showed that the statistical distance between the real distribution 
and the uniform distribution defined mathematically is 
 $\delta\sim 10^{-4}$, but a perfectly uniform random variable 
defined by the mathematics is not given. 
That is, the failure probability is $P_r({\bf K} \ne {\bf K_U})\sim 1$, and 
it is independent of $\delta$.
 
S.Takeuchi said that its result seems obvious, that is, 
 nobody can generate a perfectly uniform random variable by a physical 
process. ``Iwakoshi agreed that the experiment is physically obvious and 
that its insignificance is the point. The problem is that the QKD community 
says that when one generate $10^4$ bits key by QKD, it does not fail to 
generate a perfectly uniform random variable defined mathematically except 
for the probability $P_r({\bf K} \ne {\bf K_U}) =\delta \le {\bar{\epsilon}}
=10^{-6} \sim 10^{-14}$". 
In QKD model, since Alice and Bob cannot access the source of correlation 
controlled by Eve, the above situation has a similarity to QKD model.

Thus, the above discussions verify 
what Yuen wants to claim, and reveal the peculiarity of 
the failure probability interpretation.

\section{{\bf{Conclusions}}}
(a) In general, one needs to give an operational meaning to 
the security measure,i.e.,-trace distance.
However, one cannot interpret 
the trace distance as  ``a failure probability" of the protocol
 itself or of the indistinguishability between the generated 
key and a perfectly uniform key sequence.
In fact, the interpretation of $\epsilon$ as a failure probability is wrong 
(see figure 1).\\

(b) A one-time pad forwarded by QKD is not superior to conventional cipher 
even if such a hybrid one has information theoretic security, 
because it has a possibility that the generated key sequence 
can be instantly estimated (time is zero) by a known plaintext attack against 
the one-time pad (see figure 2). 
It is true, even if ${\bar {\epsilon} }$ is the order of $10^{-20}$, 
and further that it is impossible to realize arbitrarily small value of
 ${\bar {\epsilon}}$ due to the physical properties of quantum
 or optical channels. 
On one hand, Eve needs still time to decrypt conventional cipher. 
Thus, any phrase to the effect
 ``Information theoretic security 
 is superior to computational security" has to be carefully used when 
 one applies their own measure to security in a real-world 
setting.\\

(c) If one wants to show the usefulness of QKD, one 
has to prove the existence of a real system with $\epsilon \sim 10^{-3000}$ 
for the generated key ${\bf{|K|}}=10^4$ bits, as an example.\\
\\
Finally, users of QKD should ask QKD researchers ``why they do not show 
an understandable operational meaning of the security based on 
the quantitative value of own security measure.
A thorough discussion of QKD and a remedy to its problem
 are given in [8,17-19,20].

\section*{Acknowledgment}
I am grateful to T. Usuda and  K.Kato, and T.Iwakoshi 
for fruitful discussions.

\begin{figure}
\centering{\includegraphics[width=\columnwidth]{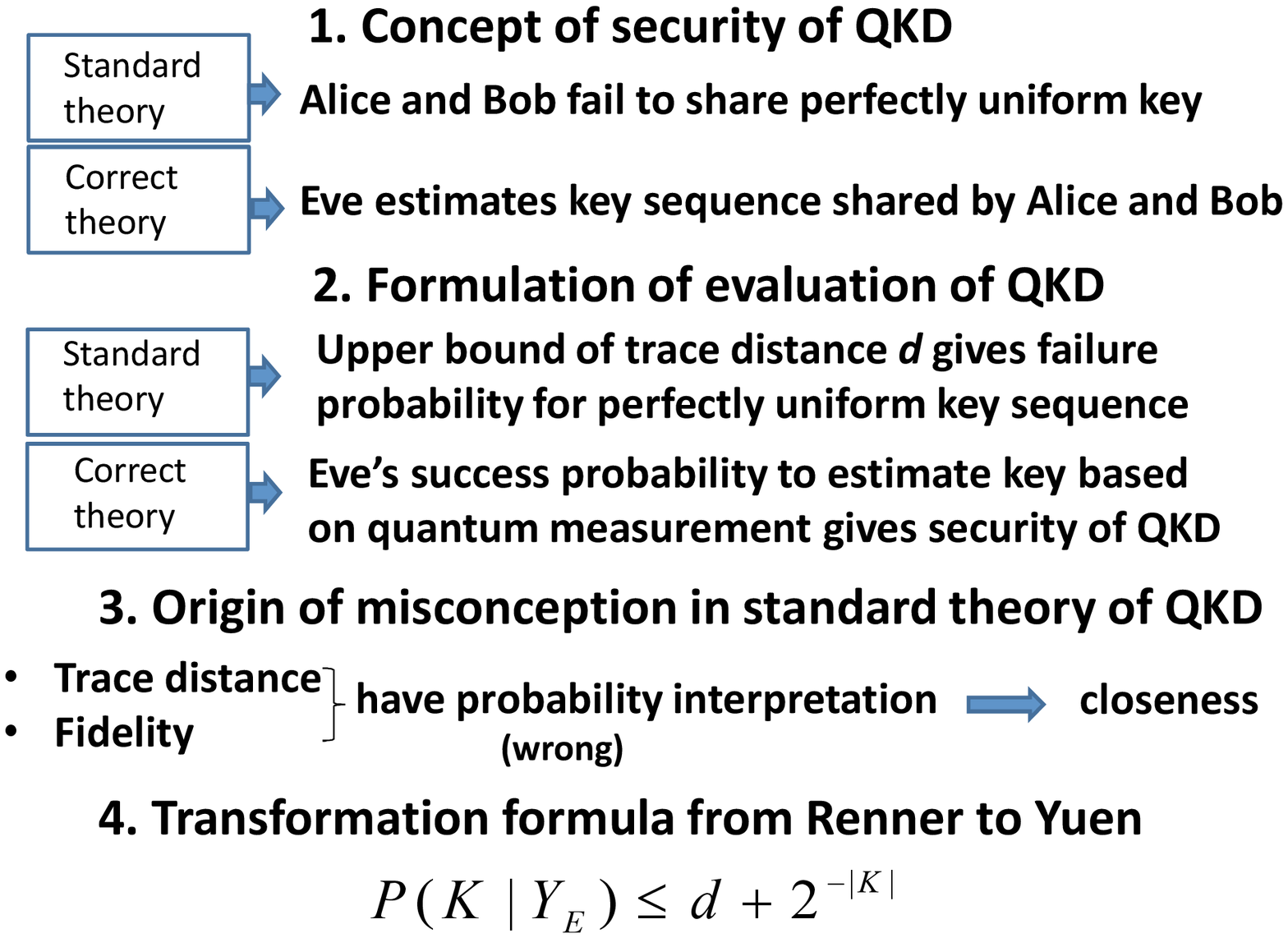}}
\caption{Difference between security formulae of QKD system. 
Standard theory of Renner is clearly summarized in [21] under 
the wrong justifications.}
\end{figure}

\begin{figure}
\centering{\includegraphics[width=\columnwidth]{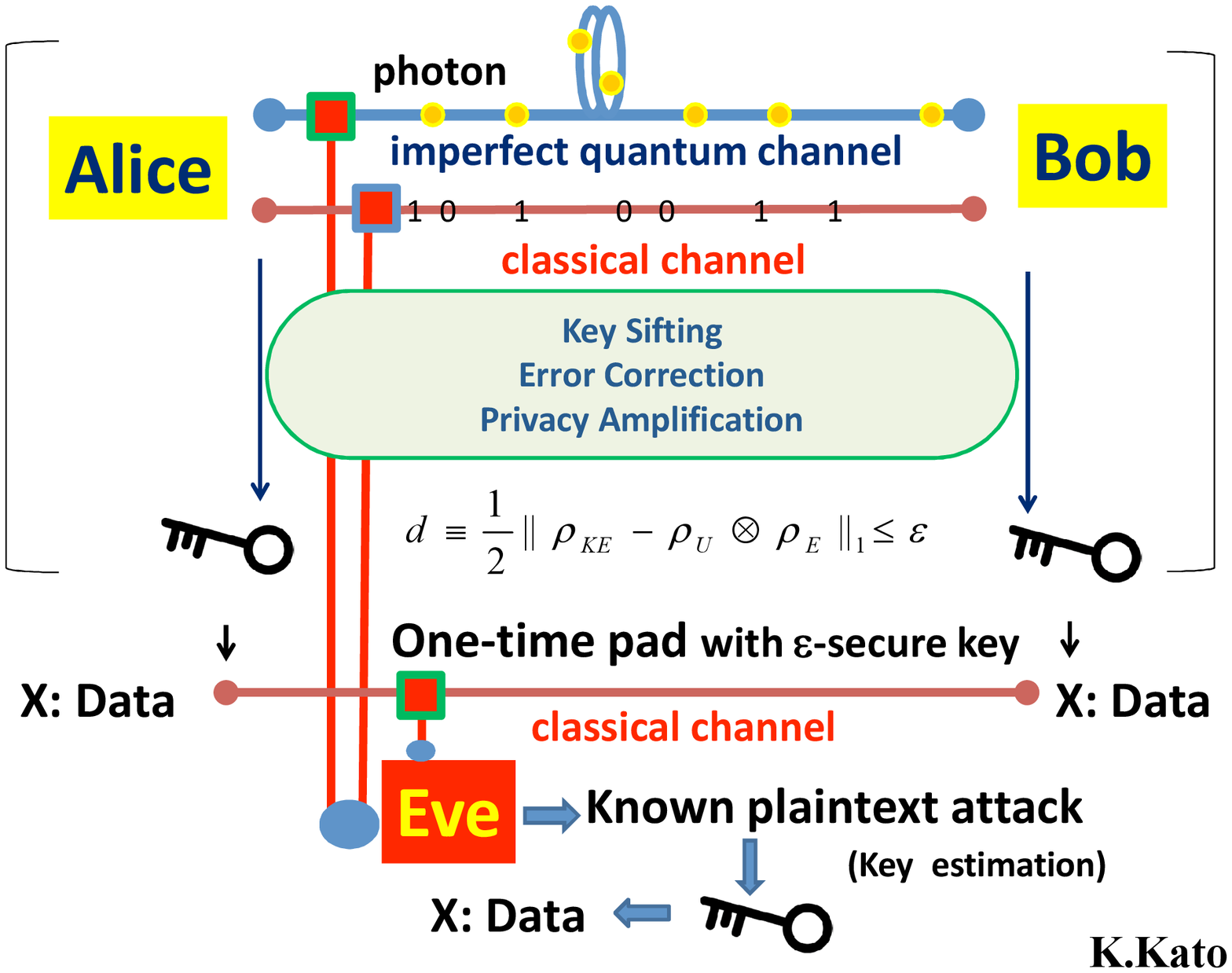}}
\caption{Schematic diagram of decryption of one-time pad with
 imperfection key. }
\end{figure}

\section*{Appendix}
\subsection{Relation between quantum detection theory and trace distance}
Portmann and Renner claimed that Helstrom's formulation gives a justification 
of the probability interpretation of the trace distance in their paper [21]. 
This is an evidence that they do not understand a physical meaning
 of  quantum detection theory. 
In fact, Yuen repeatedly explained that this justification is wrong [19], 
and Kato and Iwakoshi gave more detailed explanation.

The process of the detection has a definite physical meaning wherein 
the observer accesses to physical system described by two density operators.
The process is described by POVM $\Pi$ in general, and the minimum error in 
 the action of detection is given by Helstrom's formula 
\begin{equation}
mini_{\{\Pi\}} P_e=\frac{1}{2} (1- ||\rho_1 -\rho_2||_1)
\end{equation}
From the above result, Portmann et al interpreted that the second term itself 
 means also a probability. 
The trace distance appeared in the QKD problem does not means the action 
of the detection. It mere describes a physical situation as a closeness between 
two density operators.
The trace distance is just a parameter in detection problem, 
and trace distance itself cannot have a meaning of a probability 
in any situation.
Koashi took the same error in which he gave an interpretation 
of a probability to own fidelity formulation. His justification is such that 
the fidelity form between signal density operator and POVM gives 
a probability, so a probability interpretation of the fidelity 
is applicable to the QKD problems. But, fidelity is just inner product for 
two density operators in QKD setting. 
\subsection{Fake proof of coupling theory}
The coupling theory is a common concept in the probability theory[22].
However, Portmann and Renner have twisted the theory to justify 
own interpretation [21], using the word of ``there exist". 
Kato again clarified their trick as follows [11]:\\
In general, the inequality in coupling theory is as follows:
\begin{equation}
P(X \ne Y) \geq \delta
\end{equation}
QKD needs 
\begin{equation}
P(X \ne Y) \leq  \delta
\end{equation}
If one uses a word of "there exist", one can use $\leq$ in the coupling theory, 
because $=$ is ensured in the coupling theory.
However, there is no case of $=$ in QKD setting (see figure 3).
If Portmann et al would like to claim own justification, 
they should  discuss the coupling theory under the conditions imposed by 
physical model of QKD.
\begin{figure}
\centering{\includegraphics[width=\columnwidth]{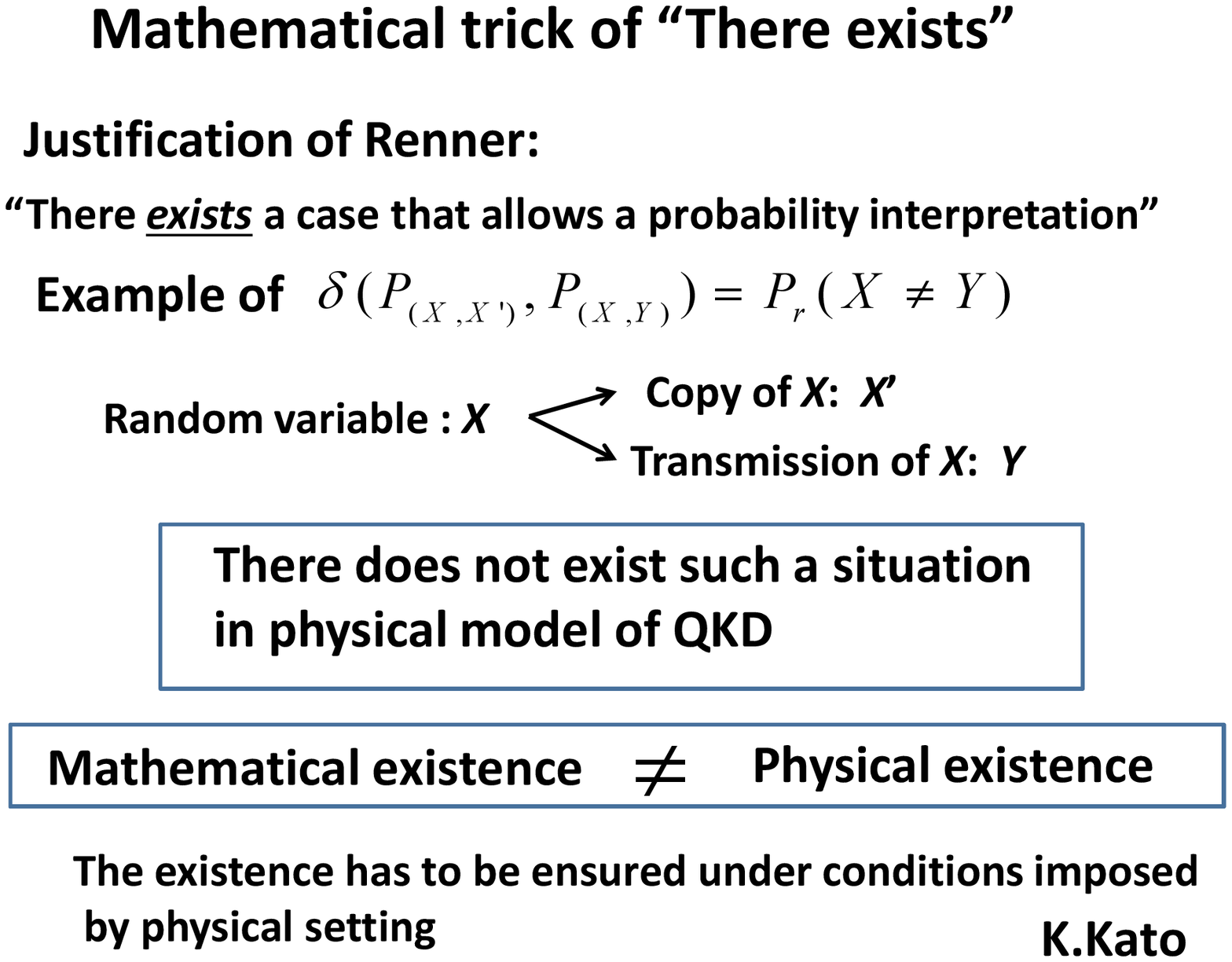}}
\caption{Origin of misconception}
\end{figure}

\subsection{Distinguishing advantage formulation vs 
 success probability formulation}
The origin of the distinguishing advantage formulation come from 
the modification of a game theory in a classical setting, wherein 
there were no clear discussion on an operational meaning of the 
distinguishability. But, the present QKD theory employed such a formulation 
without careful consideration. Thus, to ensure a cryptographic security, 
the distinguishing advantage formulation is forced to employ a failure 
probability interpretation of trace distance. However there are two intrinsic 
problems.\\
(1) Eve has ``knowledge" on key before privacy amplification (PA). 
But PA can only handle a subset of Eve's whole knowledge.
Although the output of PA seems a uniform key, it does not mean 
that the key is also uniform key for Eve.\\
(2) The trace distance for the final key sequence also has no binary
 interpretation like ``yes or no" due to  failure probability.\\
 Why can one say based on such features of the formalism that 
the trace distance ensures the composability.
Thus there is a chain of misconception in this formulation.
One has to employ success (or guessing) probability formulation based 
on M-ary quantum detection theory which have been developed by 
Holevo, Yuen and Hirota school 
to unify information theoretic security. Fortunately, an evaluation of 
the trace distance can be transformed into success probability evaluation 
by Yuen's formula.
\begin{equation}
P({\bf{K}}|{\bf{Y}}_E)  \leq d_{QKD} + 2^{-|{\bf{K}}|}
\end{equation}

\subsection{Total efficiency of key generation}
A communication performance is one of the most important features
 in information science. We cannot avoid an imperfection of 
communication equipments in a real world. Especially, the energy loss 
in a channel and a detector for single photon gives serious degradation of 
communication performance. In the conventional QKD system, 
random bits of about 50 Gbit/sec ($50\times10^9$) are transmitted from 
Alice, and the final keys of 300 Kbit/sec ($300\times 10^3$) are shared  
 between Alice and Bob.
Thus to realize a one-time pad communication system with 300 Kbit/sec,
QKD system uses an optical communication technology for 50 Gbit/sec.
So far there was no such a greatly inefficient technology.

\end{document}